\documentclass[letterpaper,twocolumn,showpacs,superscriptaddress,amsmath,amssymb,prb]{revtex4}

\usepackage{graphicx,color}
\usepackage[colorlinks,plainpages=false,linkcolor=black,urlcolor=blue,citecolor=blue,pdfpagemode=UseNone,pdfstartview=FitH]{hyperref}

\begin{document}

\title{Anomalously enhanced photoemission from the Dirac point and other peculiarities in the self-energy of the surface state quasiparticles in Bi$_2$Se$_3$}

\author{A.~A.~Kordyuk}
\affiliation{Institute for Solid State Research, IFW-Dresden, P.O.Box 270116, D-01171 Dresden, Germany}
\affiliation{Institute of Metal Physics of National Academy of Sciences of Ukraine, 03142 Kyiv,  Ukraine}

\author{V.~B.~Zabolotnyy}
\author{D.~V.~Evtushinsky}
\author{T.~K.~Kim}
\author{B.~B\"{u}chner}
\affiliation{Institute for Solid State Research, IFW-Dresden, P.O.Box 270116, D-01171 Dresden, Germany}

\author{\\I.~V.~Plyushchay}
\affiliation{Taras Shevchenko National University of Kyiv, 01601 Kyiv, Ukraine}

\author{H.~Berger}
\affiliation{Institute of Physics of Complex Matter, EPFL, 1015 Lausanne, Switzerland}

\author{S.~V.~Borisenko}
\affiliation{Institute for Solid State Research, IFW-Dresden, P.O.Box 270116, D-01171 Dresden, Germany}

\begin{abstract}
Line shape analysis of the photoemission intensity from the surface states of Bi$_2$Se$_3$ reveals two unusual features: spectral line asymmetry and anomalously enhanced photoemission from the Dirac point. The former can be described by the one-particle spectral function assuming that the self-energy has an energy-momentum dependent contribution to the imaginary part, which changes sign when crossing both the dispersion curves and the energy of the Dirac point. The anomalous enhancement can hardly be understood as a self-energy effect, while a final state interference seems to be a more plausible explanation.
\end{abstract}

\maketitle

\section{Introduction}

The topological insulators, as a new kind of matter, and emerging new physics have been predicted theoretically \cite{Fu} and later observed in a number of the state-of-the-art experiments. Angle resolved photoemission spectroscopy (ARPES), applied to the topological insulators \cite{Fu,Moore}, has been playing a major role in their promotion and study \cite{Xia,Hsieh,Chen}. The experimentally observed cone-like dispersion of the surface states with the spin degenerate Dirac point is now considered as a hallmark of the topological surface states \cite{Moore}. In this sense, the compound Bi$_2$Se$_3$ with a single Dirac cone in the Brillouin zone \cite{Xia,Hsieh} provides the most elementary case for the in-depth ARPES study of those states. Analyzing the spectral weight distribution in spectra of Bi$_2$Se$_3$, we observe deviations from the expected one particle spectral function: an anomalously enhanced photoemission from the vicinity of the Dirac point and unusual asymmetries of the surface states spectral function. Having excluded the possible experimental artifacts and trivial explanations, we show that the $k\omega$-dependent component to the self-energy changes sign while crossing both the dispersion curves and the energy of the Dirac point. We also conclude that the unexpected intensity enhancement at the Dirac point is hard to describe by the single particle spectral function. Possible mechanisms for the enhancements are discussed.

\begin{figure}[t]
\begin{center}
\includegraphics[width=0.28\textwidth]{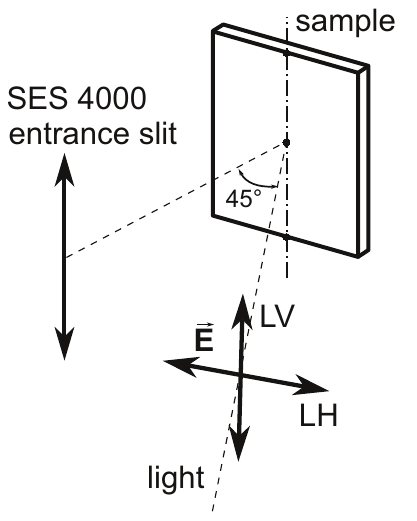}
\caption{
The geometry of our ARPES experiment: LH and LV stay for the linear horizontal and linear vertical polarizations of the incident light.
\label{ARPES}}
\end{center}
\end{figure}

\begin{figure*}
\begin{center}
\includegraphics[width=0.96\textwidth]{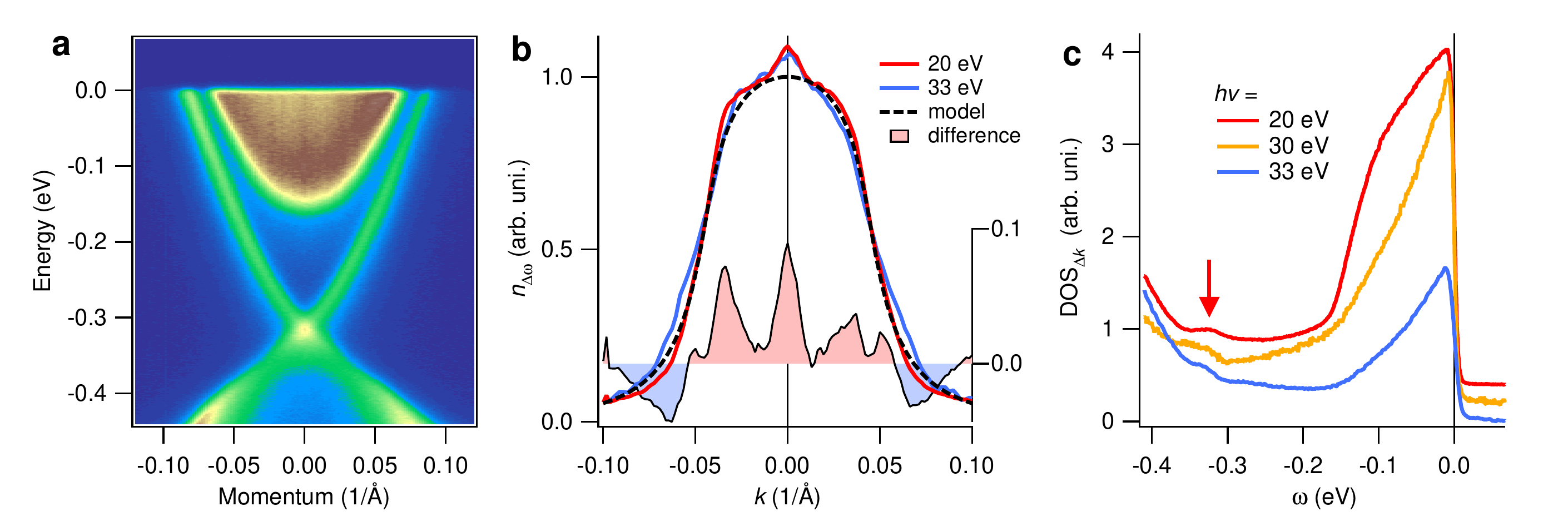}
\caption{
ARPES image of the Dirac cone in Bi$_2$Se$_3$ measured close to the $\Gamma$K direction, $T$ = 20\,K (a). Energy- (b) and momentum- (c) integrated ARPES spectra represent partial $n(k)$ and partial DOS, respectively. The difference between 20 eV experimental data and the model, shown in panel (b), refers to the right axis. The arrow in panel (c) marks the position of the Dirac point.
\label{data1}}
\end{center}
\end{figure*}

The paper is organized as follows. In Section \ref{s_exp} we describe the experiment and the results, pointing out on the anomalously enhanced photoemission from the Dirac point that can be seen in the raw data. In Section \ref{s_analysis}, performing a deeper data analysis, we: (i) explain why the observed enhancement is unexpected based on one particle spectral function and the related sum rules (Section \ref{s_sumrule}); (ii) exclude possible artifacts (\ref{s_artif}); (ii) present detailed comparison of the experimental data and simple models (\ref{s_fit}); (iv) detect an unusual spectral line asymmetry and show that it can be described by one-particle spectral function with a certain dependence of the self-energy on energy and momentum (\ref{s_selfen}), and (v) demonstrate that the anomalous enhancement cannot be understood if a physically viable model for the self energy is considered (\ref{s_selfen}). An alternative explanation for it, a final state interference, is discussed in Section \ref{s_interfe}.

\section{Experiment and Results}
\label{s_exp}
Single crystals of Bi$_2$Se$_3$ were grown using the Bridgman method starting with high-purity Bi and Se elements. ARPES experiments have been performed at the ``1$^3$" and ``1$^2$" beamlines equipped with SES 4000 and SES 8000 analyzers at the synchrotron radiation source BESSY II. Fig.\,\ref{ARPES} shows the geometry of our ARPES experiment, in which the entrance slit of the electron analyzer is fixed, but the sample can be translationally moved and rotated. More details can be found in Ref.\,\onlinecite{TopoGating}. The data have been collected at different temperatures in the range from 1\,K to 30\,K revealing independence of the observed anomalies on the temperature.

In Fig.\,\ref{data1}a, we show the aforementioned ``hallmark" of the topological insulators---a typical ARPES spectrum of Bi$_2$Se$_3$ measured close to the $\Gamma$K direction with the surface state dispersion converging at the Dirac point in the center of the Brillouin zone. In contrast to the three-dimensional bulk states, which fill here the parabola-enclosed region at the top of the spectrum, the surface states are truly two-dimensional, which makes them a perfect object for an accurate ARPES study of the parameters the spectral function depends on. However, we have found that the deviations of the photoemission intensity from the expected spectral function of 2D electronic excitations are essential and require thorough investigation.

The most striking effect appears as an anomalously enhanced photoemission from the Dirac point. This anomaly is clearly seen in Fig.\,\ref{data1}b as a small sharp peak with the full width at half maximum (FWHM) about 0.012\,\AA$^{-1}$ on a partial momentum distribution function, $n_{\Delta\omega}(k)$, that is defined by integration of the corresponding ARPES spectrum in the energy range from $-0.4$ eV to $-0.2$ eV. The range is chosen to be narrow enough to get rid of the influence of the bulk states, but wide enough to expect a plateau around the center of the Brillouin zone ($\pm 0.04$\,\AA$^{-1}$). The local enhancement can also be seen (Fig.\,\ref{data1}c) on partial density of states, DOS$_{\Delta k}(\omega)$, that is defined by integration in the momentum window $\pm 0.12$\,\AA$^{-1}$. In order to understand why the enhancement is anomalous, one needs to consider the structure of the spectral function and its relation to the ARPES spectrum.

\section{Data analysis and discussion}
\label{s_analysis}
\subsection{Trivial spectral function}
\label{s_sumrule}
The photoemission intensity from 2D or quasi-2D states are expected to be well described by the product of the Fermi function, $f(\omega)$, and the one-particle spectral function, $A(\omega,\mathbf{k})$ (a function of energy and in-plane momentum). For the photoemission current from the $n$-th band, neglecting the experimental resolution and photoemission background, one can write
\begin{eqnarray}
I_n(\omega,\mathbf{k}) &=& M(n,h\nu,\omega,\mathbf{k}) f(\omega) A_n(\omega,\mathbf{k}),
\label{In}
\end{eqnarray}
where the transition probability (the squared modulus of photoemission matrix elements) and some extrinsic factors (such as detector efficiency) are included in a factor $M$, and $A_n$ is supposed to have a simple representation in terms of bare electron dispersion, $\varepsilon_\mathbf{k}^n$ of the $n$-th band, and the corresponding one-particle self-energy, $\Sigma_n = \Sigma'_n + i\Sigma''_n$:
\begin{eqnarray}
A_n(\omega,\mathbf{k}) &=& \frac{1}{\pi} \frac{-\Sigma''_n(\omega,\mathbf{k})}{(\omega - \varepsilon_\mathbf{k}^n - \Sigma'_n(\omega,\mathbf{k}))^2+\Sigma''_n(\omega,\mathbf{k})^2}.
\label{An}
\end{eqnarray}

For the sake of simplicity, we will consider the spectrum such as shown in Fig.\,\ref{data1}a, i.e.\;measured in a cut of the $\mathbf{k}$-space through the center of the Brillouin zone, as one-dimensional, i.e.\;$\mathbf{k} = (k_x,0) = (k,0)$. In Appendix\;\ref{AppA} we show that the finite momentum resolution along $k_y$ does not affect the eligibility of this approximation. Then, the spectrum can be considered as a sum of intensities from two bands, $I_1(\varepsilon_k^1) + I_2(\varepsilon_k^2)$, the states in which can be enumerated in a way to trace smooth monotonic dispersions $\varepsilon_{k}^1 = \varepsilon_{-k}^2 = \varepsilon_k$. So, one can write
\begin{eqnarray}
I(\omega,k) = I_1(\omega,k) + I_1(\omega,-k).
\label{Isum}
\end{eqnarray}
This means that one does not expect to see any sharp features in energy or momentum integrated $I(\omega,k)$ except those present in similarly integrated $I_1(\omega,k)$.

The dependence of the coefficient $M$ on $\mathbf{k}$ and $\omega$ in Eq.\,(\ref{In}) is determined by the photoemission matrix elements and can be neglected in our analysis. Indeed, the enhancement is localized within less than 0.05 eV, which is by an order of magnitude smaller than one would expect from the variation of matrix elements with energy \cite{ME}. In addition, we have observed the effect at several excitation energies, as shown in Fig.\,\ref{data1}b,c. Both observations practically rule out $M(\omega,\mathbf{k})$ as a source of the anomaly reported here and, in the following, we focus on the one-particle spectral function and examine whether the observed enhancement can be described by a sophisticated $\omega \mathbf{k}$-dependent self-energy or should be caused by an extrinsic mechanism.

\subsection{ARPES sum rules}
\label{s_sumrules}
Now, let us remind the sum rules that the one-particle spectral function must comply with.\cite{Randeria} The simplest one comes from its definition and says that its integral over all frequencies is unity:
\begin{eqnarray}
\int_{-\infty}^{\infty}{A(\omega,\mathbf{k}) d\omega} = 1.
\label{sumA}
\end{eqnarray}
A more practical one,
\begin{eqnarray}
\int_{-\infty}^{\infty}{A(\omega,\mathbf{k})f(\omega) d\omega} = n(\mathbf{k}),
\label{sumAf}
\end{eqnarray}
gives the momentum distribution function $n(\mathbf{k})$ that is close to 1 when $A(\omega)$ is peaked well below the Fermi-level, i.e.\;when $(\varepsilon_\mathbf{k} + \Sigma')^2 \gg \Sigma''^2$ for $\varepsilon_\mathbf{k} + \Sigma' < 0$.

A similar function, which can be extracted from ARPES spectra, we will call a ``partial'' momentum distribution function,
\begin{eqnarray}
n_{\Delta \omega}(\mathbf{k}) = \int_{\omega_1}^{\omega_2}{A(\omega,\mathbf{k})f(\omega) d\omega},
\label{sumAf}
\end{eqnarray}
where $\Delta \omega = \omega_2 - \omega_1$ is the range of integration that is always finite in experiment, but in this case we narrow it deliberately to avoid the influence of the electronic bands that are out of interest.

The difference between $n(\mathbf{k})$ and $n_{\Delta\omega}(\mathbf{k})$ is due to the tails of $A(\omega)$ that are cut by the window of integration. Since, within the assumptions made above, the profile shown in Fig.\,\ref{data1}b represents $n_{\Delta\omega}(\mathbf{k})$, one can expect that the observed peak might be associated with a sharp narrowing of the energy distribution curves (EDC) at $k = 0$ that, in turn, corresponds to a sharp minimum in $\Sigma''(\mathbf{k})$ at this point.

Similarly, while the summation over all $\mathbf{k}$-states gives the density of occupied electronic states
\begin{eqnarray}
\mathrm{DOS}(\omega) = \sum_{\mathbf{k}}{A(\omega,\mathbf{k})f(\omega)},
\label{DOS}
\end{eqnarray}
it is often useful\cite{VallaSci99,KordyukPRB09} to define a partial density of states
\begin{eqnarray}
\mathrm{DOS}_{\Delta k}(\omega) = \int{A(\omega,k)f(\omega) d\mathbf{k}},
\label{pDOS}
\end{eqnarray}
which can be compared to the experimentally derived functions such as shown in Fig.\,\ref{data1}b. In our case, the region of integration is determined by the momentum window $\Delta k$ from $-0.12$\,\AA$^{-1}$ to 0.12\,\AA$^{-1}$ along the slit of the analyzer and by the momentum resolution perpendicular to the slit, the effect of which is considered in Appendix \ref{AppA}.

One can see from Eq.\,(\ref{An}) that both DOS$(\omega)$ and DOS$_{\Delta k}(\omega)$ are roughly constant when $\varepsilon_k$ is close to linear and $\Sigma$ can be considered as weakly $\mathbf{k}$-dependent. This has been shown to be the case for a number of materials including strongly correlated cuprates.\cite{VallaSci99,SE} Thus, again, the observed enhancement of the photoemission signal seen in Fig.\,\ref{data1}c at  the Dirac point energy might be caused by a sharp narrowing of the momentum distribution curves (MDC), which is a consequence of a sharp minimum in $\Sigma''(\omega)$ at this energy.

In Section \ref{s_selfen} we examine whether it is possible to describe simultaneously the peaks on both $n_{\Delta\omega}(\mathbf{k})$ and DOS$_{\Delta k}(\omega)$ functions with a $\omega\mathbf{k}$-dependent self-energy, but before, we have to prove that the effect is real, excluding possible experimental artifacts.

\subsection{Excluding trivial artifacts}
\label{s_artif}
First, one may question whether such an effect can be a result of an inaccurate position in $\mathbf{k}$-space or of a finite momentum resolution. Both cases are similar to the case of a finite gap due to lifting of the degeneracy at the Dirac point. Any of these reasons would result in a dip rather than a peak in DOS$_{\Delta k}(\omega)$ and should not affect $n_{\Delta\omega}(k)$ at all (see Appendix \ref{AppA}) which is in contrast to our observation.

Second, one should examine a possible non-linearity of the detector, which has earlier been reported to be an issue with similar analyzers \cite{Kay,Mannella,Wicks}. Indeed, the observed enhancement can be qualitatively understood if the readout intensity as a function of actual photocurrent increases faster than linearly. One can notice, however, that it is an unlikely reason for the observed anomaly since its width in momentum is very narrow, less than a half of the minimum line width that can be found in the spectrum. Nevertheless, in Appendix \ref{AppB}, by careful analysis and modeling the ARPES spectrum, we show that the actual non-linearity of the detecting system we use cannot result in the observed anomaly.

\begin{figure*}
\begin{center}
\includegraphics[width=\textwidth]{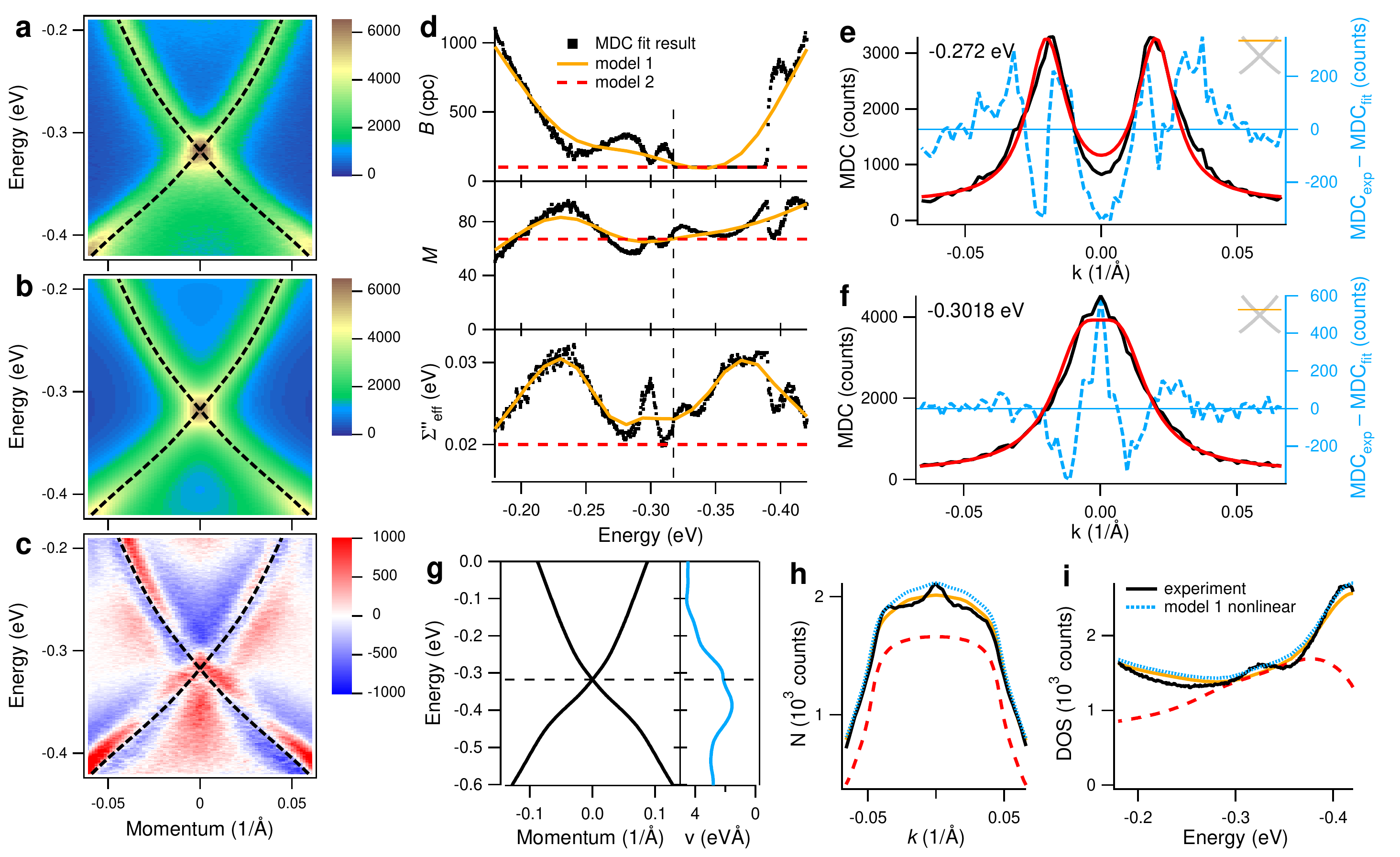}
\caption{
The experimental data (a), the result of the fit-based model (model 1) (b), and their difference (c), superimposed by the MDC dispersion (dashed lines). The spectral function parameters derived from the MDC fit, as described in the text (d): $\Sigma''$ is an effective imaginary part of the self-energy, $M = v M_\omega$, where $M_\omega$ is the MDC area and $v$ is the bare band velocity, $B$ is the momentum independent background in counts per channel. Examples of the MDC fit at 0.275 eV (e) and 0.302 eV (f) binding energies: experimental data (black symbols), two-lorentzian fit (red solid line), and their difference (dashed blue line). The experimentally derived band dispersions and the band velocity (g). Partial momentum distribution $n_{\Delta\omega}(\mathbf{k})$ (h) and partial DOS (i) derived from experiment (solid black line) and the model 1 and 2 as well as for the case of 10\% nonlinearity applied to the model 1.
\label{datafit}}
\end{center}
\end{figure*}

\subsection{Fit to the model}
\label{s_fit}
In order to better reveal the differences between the measured spectrum and the expected model further, we consider the spectrum as a set of MDC's \cite{VallaSci99} and fit them to
\begin{eqnarray}
I(\omega,k) = I_1(\omega,k) + I_1(\omega,-k) + B(\omega),
\label{I}
\end{eqnarray}
rewriting $I_1(\omega,k)$ in an MDC form:

\begin{eqnarray}
I_1(\omega,k) &\approx& \frac{M_{\omega} f(\omega)}{\pi} \frac{W(\omega)}{[k_m(\omega) - k]^2+W^2(\omega)}.
\label{I1}
\end{eqnarray}
Here $M_\omega = M/v$; the MDC width (half width at half maximum) $W(\omega) = -\Sigma''(\omega)/v(\omega)$; the position of MDC maximum $k_m(\omega) = [\omega - \varepsilon_0(\omega) - \Sigma'(\omega)]/v(\omega)$; $\varepsilon_0 = \varepsilon_k(k_m) - v(\omega)k_m$; $v(\omega) \approx (dk_m/d\omega)^{-1}$; $B(\omega)$ stays for both the momentum independent background and contribution from the bulk states.

In Fig.\,\ref{datafit} we shows the results of such a fit and compare the experimental spectrum (a) to the model (b). Panel (c) represents their difference. The two Lorentzian fit to the experimental MDC is stable everywhere except $\pm$\,5\,meV around the Dirac point. So, in the first step we derive $k_m(\omega)$ in the fit stability range, interpolate it in the vicinity to the Dirac point and smooth it with the 10 meV smoothing window. The result is show in Fig.\,\ref{datafit}g. Then, holding $k_m(\omega)$, we derive other parameters, using the only $B > 0$ constrain, and put in the model (model 1) the smoothed functions $\Sigma''(\omega)$, $M(\omega)$, and $B(\omega)$, as shown in Fig.\,\ref{datafit}d. By the smoothing, we filter out the rapid variation of the fitting parameters in the close vicinity to the Dirac point.

A one-to-one comparison of the experimental and model MDCs shows that the experimental spectrum differs from the expected one in two ways. First, an unexpected MDC lineshape is observed in the vicinity to the Dirac point: a narrow peak has appeared in a middle of two nearly merged peaks, as shown in Fig.\,\ref{datafit}f. Evidently, it is this peculiarity that is responsible for the aforementioned enhancement of the photoemission intensity in the energy- and momentum-integrated spectra. Second, there are systematic deviations of the experimental MDCs from the line shape expected for the spectral function with the momentum independent self-energy, as shown in Fig.\,\ref{datafit}e. Those deviations can be understood as an asymmetry of each MDC peak in respect to the position of its maximum. Fig.\,\ref{datafit}c shows that these deviations are systematic. In the following sections we discuss the both anomalies separately, but first, having in hands the developed fit-based model, we address the question of detector linearity more carefully.

Assuming a weak non-linearity of the the read-out count rate on real photoemission intensity $D = I(1+\alpha I/I_{max})$, we can show that for the detector and count rates we use, $\alpha < 0$  (see Appendix \ref{AppB}), i.e.\;one would expect a depletion rather than enhancement of the intensity in the Dirac point. Nevertheless, we model the case of $\alpha =0.1$, recently reported for similar detectors \cite{Wicks}. In Fig.\,\ref{datafit}(h,i), one can see that although the effect of 10\% non-linearity is noticeable, it cannot produce such sharp peaks as observed in $n_{\Delta\omega}(k)$ and DOS$_{\Delta k}(\omega)$.

\begin{figure}[t]
\begin{center}
\includegraphics[width=0.42\textwidth]{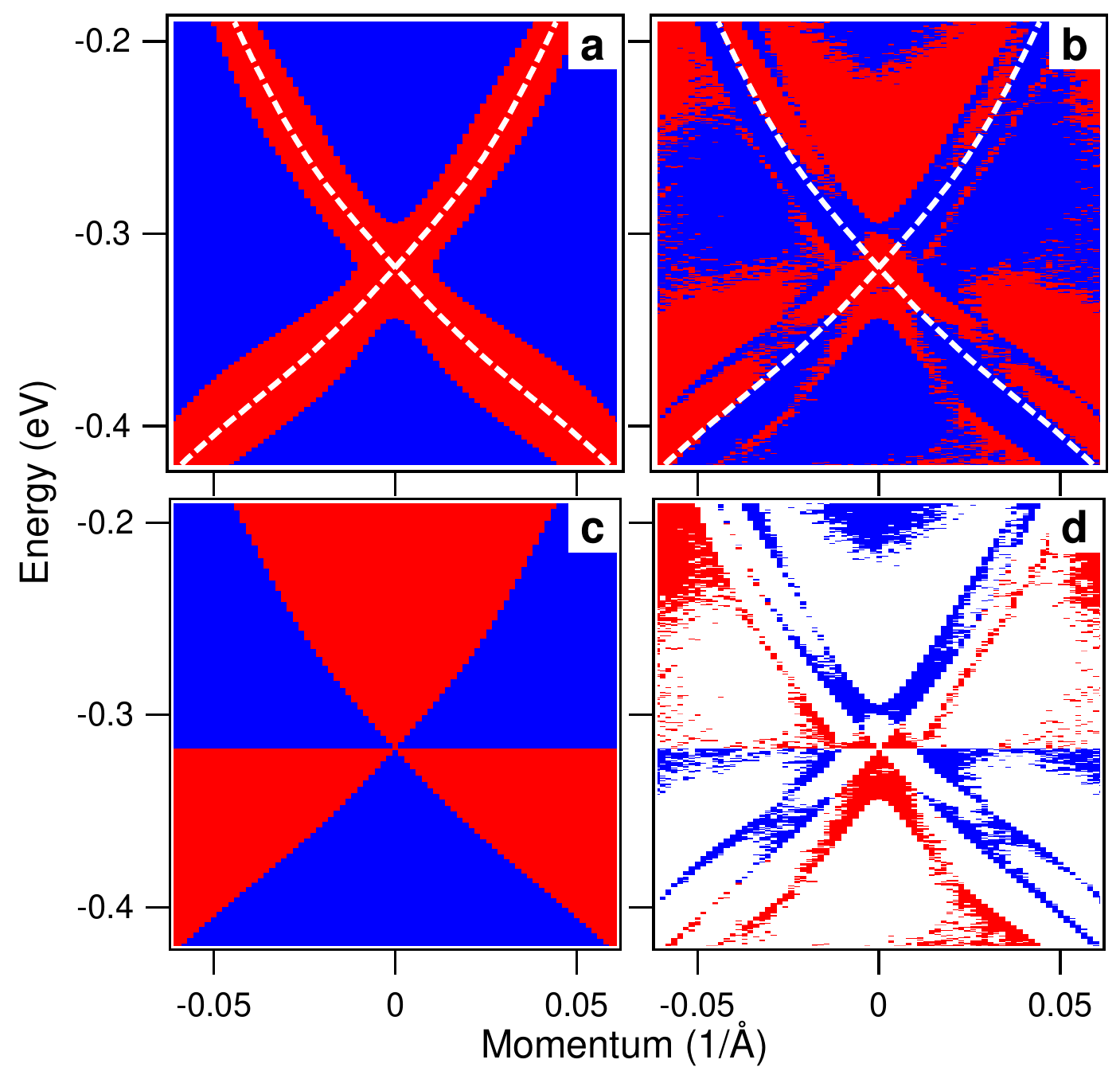}
\caption{
A pattern of the energy dependent part of the self-energy: (a) sign of $S(\omega,k)$; (b) sign of $\sigma''(\omega,k)$, (c) suggested symmetry (center), and (d) their difference; positive is red, negative is blue, zero is white.
\label{symm}}
\end{center}
\end{figure}

\subsection{Line shape asymmetry}
\label{s_selfen}

Let us assume that the difference between the real and model spectral functions, such as shown in Fig.\,\ref{datafit}c, is the result of a small component of the self-energy with strong dependence on energy and momentum, and find its simplest form that can explain both the spectral line asymmetry and the anomalously enhanced photoemission from the Dirac point.

We start with the asymmetry issue and write the total self-energy as $\Sigma(\omega,\mathbf{k}) = \Sigma_0[1+\sigma(\omega,\mathbf{k})]$, where $\Sigma_0$ is a main self-energy part that may be constant or may vary slowly with $\omega$. For $\sigma'' \ll 1$, completely neglecting a contribution of $\sigma'$ to the spectral function (that violates the Krames-Kronig consistency\cite{SE} but does not affect our finite result), one can derive:
\begin{eqnarray}
\Delta A(\omega, k) = A(\Sigma)-A(\Sigma_0) \simeq S \sigma'',
\label{dA}
\end{eqnarray}
where
\begin{eqnarray}
S = -\frac{1}{\pi}\left[\frac{\xi_k^{2}-\Sigma''^{2}_0}{(\xi_k^{2}+\Sigma''^{2}_0)^2} + \frac{\xi_{-k}^2-\Sigma''^{2}_0}{(\xi_{-k}^2+\Sigma''^{2}_0)^2} \right] \Sigma''_0,
\label{S_factor}
\end{eqnarray}
and $\xi_k = \omega - \varepsilon_k - \Sigma'_0$. Far away from the Dirac point, i.e., for $|\xi_{k}-\xi_{-k}| \gg 2\Sigma''_0$, and close to the dispersion, e.g., for $\xi_{k} \ll \Sigma''_0$,
\begin{eqnarray}
\Delta A(\omega, k) = \sigma''(\omega, k)/\pi \Sigma''_0.
\end{eqnarray}

From Eq.\,(\ref{dA}) one can see that $\Delta A(\omega, k)$ changes sign when either $S(\omega,k)$ or $\sigma''(\omega,k)$ does. The former happens when the spectral function intensity is about half of its maximum, see Fig.\,\ref{symm}a or Eq.\,(\ref{S_factor}), but the sign of $\Delta A(\omega, k)$ changes also at $\xi_{k} = 0$, $\xi_{-k} = 0$, and $\omega = \omega_{\mathrm{DP}}$, as it follows from Fig.\,\ref{datafit}c or Fig.\,\ref{symm}b. One can conclude therefore that $\sigma''$ should change the sign when crossing either the dispersion curves or the energy of the Dirac point, as it is shown in Fig.\,\ref{symm}c, which can be a signature of a spin-flip scattering \cite{Yang}.

Note that while the sign-change pattern of $\sigma''$ shown in Fig.\,\ref{symm}c can be described by its dependence on $\omega$ only, the dependence on $k$ is still necessary to explain the asymmetry of the MDC peaks.

\subsection{Anomalous peak at Dirac point}
\label{s_peak}

\begin{figure}[t]
\includegraphics[width=0.47\textwidth]{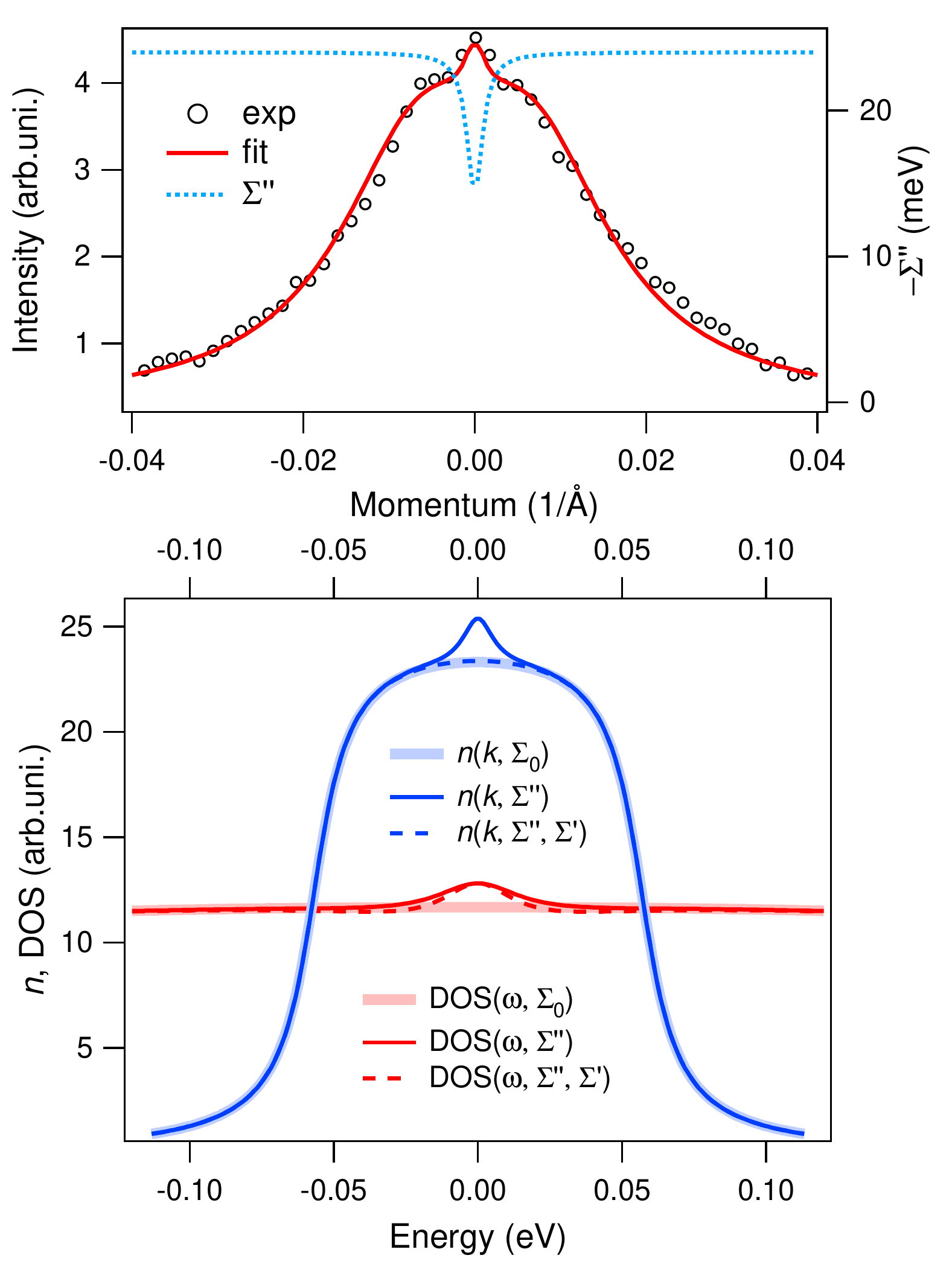}
\caption{
Momentum dependent self-energy. The peaked MDC can be a consequence of a sharp decrease of the scattering rate at $k=0$ (top). Both $n_{\Delta\omega}(k)$ and DOS$_{\Delta k}(\omega)$ can only be fitted to the experiment if the Kramers-Kronig consistency is not taken into account  (bottom).
\label{kdep}}
\end{figure}

One can certainly describe a single MDC with a momentum dependent self-energy. The sharp peak that appears in the middle of the MDCs in the vicinity to the Dirac point, as shown in Fig.\,\ref{datafit}f, would require the scattering rate to have a sharp drop at $k = 0$. In Fig.\,\ref{kdep} (top), we fit the model MDC, defined by Eq.\,(\ref{I}), to the experimental one writing the imaginary part of the self-energy as $\Sigma''(k) = \Sigma_0''[1+\delta''(k)]$, where $\delta''(k) = \alpha w_k^2/(k^2+w_k^2)$, with $\Sigma_0'' = -0.024$ eV, $\alpha = -0.4$, $w_k = 0.0011$ {\AA}$^{-1}$, $k_m = 0.0068$ {\AA}$^{-1}$, and the band velocity at the Dirac point $v = 2.12$ eV{\AA}.

However, further analysis shows that these parameters do not describe the whole spectrum and, moreover, that the dependence of the self-energy on $k$ only cannot describe both $n_{\Delta\omega}(k)$ and DOS$_{\Delta k}(\omega)$ simultaneously. In particular, fitting the model to experimental $n_{\Delta\omega}(k)$ gives very different $\Sigma_0'' = -0.02$ eV, $\alpha = -0.7$, $w_k = 0.008$, and, in addition, results in three times higher peak in the modeled DOS$_{\Delta k}(\omega)$ than it is observed experimentally.

One may hope that the problem can be resolved if $\delta''$ depends on both $k$ and $\omega$, but we have found that it cannot. Despite the linear dispersion, the influence of the momentum and energy dependences of the self-energy on $n_{\Delta\omega}(k)$ and DOS$_{\Delta k}(\omega)$ is not symmetric. The reason for this is in the Kramers-Kronig relation between energy dependences of the real and imaginary parts of the self-energy. One can show that if the Kramers-Kronig consistency is taken into account, the energy dependence of $\Sigma''$ has virtually no effect on $n_{\Delta\omega}(k)$. To illustrate this, we model the depletion of $\Sigma''$ as
\begin{eqnarray}
\delta''(\omega, k) = \frac{\alpha}{\left(\omega/w_\omega\right)^2 + \left(k/w_k\right)^2 + 1}.
\end{eqnarray}
This assumes the Kramers-Kronig consistent counterpart
\begin{eqnarray}
\delta'(\omega, k) = \frac{-\omega/w_\omega}{\sqrt{\left(k/w_k\right)^2 + 1}}\frac{\alpha}{\left(\omega/w_\omega\right)^2 + \left(k/w_k\right)^2 + 1}
\end{eqnarray}
contributes to the real part of the self-energy as $\Sigma'(\omega,k) = \Sigma_0''\delta'(\omega, k)$.

In Fig.\,\ref{kdep} (bottom) we show that both $n_{\Delta\omega}(k)$ and DOS$_{\Delta k}(\omega)$ can be fitted simultaneously only in an artificial case when $\delta'$ is set to zero. This case is presented by the solid curves which are modeled to fit to the corresponding experimental $n_{\Delta\omega}(k)$ and DOS$_{\Delta k}(\omega)$ with the following parameters: $\Sigma_0'' = -0.02$ eV, $\alpha = -0.3$, $w_\omega = 0.04$ eV,  $w_k = 0.008$ {\AA}$^{-1}$. One can see (dashed curves) that taking into account the Kramers-Kronig consistency essentially eliminates the peak in $n_{\Delta\omega}(k)$. So, the energy dependence of $\Sigma''$ cannot help to reconcile the discussed anomaly in the experimental spectrum with the one particle spectral function. This calls for more rigorous consideration of the photoemission process.

\subsection{Final state interference}
\label{s_interfe}
A plausible explanation for the observed enhancement and interference-like spectrum might be a finite state interference of the photoelectrons emitted from the $\mathbf{k}_\omega$ and $-\mathbf{k}_\omega$ states in the vicinity to the Dirac point.

Let us assume that (1) the width of a single MDC is defined by the size, $d$, of the region from where the photoelectrons are emitted coherently, and (2) the photoemission from $\mathbf{k}_\omega$ and $-\mathbf{k}_\omega$ states from this region remains also coherent. For the sake of simplicity, the former can be described by the Fraunhofer diffraction and for the photoemission intensity from a single $k_\omega$ state one can write
\begin{eqnarray}
I_1(\alpha,\theta) \propto \left[ \frac{\sin[\xi \sin(\alpha + \theta)]}{\xi \sin(\alpha + \theta)} \right]^2,
\end{eqnarray}
where $\alpha$ is the angle of electron emission, $\theta$ is the angle which corresponds to the initial state of electron, such as $\hbar k_\omega = \sqrt{2m E_k} \sin(\theta)$, $\xi = \pi d/\lambda$, $E_k$ is the kinetic energy of photoelectron, $\lambda = 2\pi \hbar/\sqrt{2m E_k}$ is its wave length, and $m$ is the electron mass.

Then, based on the second assumption and for small $\alpha$ and $\theta$, the photocurrent from $k_\omega$ and $-k_\omega$ states can be written, instead of Eq.\,(\ref{Isum}), as
\begin{eqnarray}
I(\alpha,\theta) \propto \left[ \frac{\sin[\xi (\alpha + \theta)]}{\xi (\alpha + \theta)} + \frac{\sin[\xi (\alpha - \theta)]}{\xi (\alpha - \theta)}\right]^2,
\end{eqnarray}
that is similar to the famous two-slit diffraction pattern.

\begin{figure}[t]
\begin{center}
\includegraphics[width=0.48\textwidth]{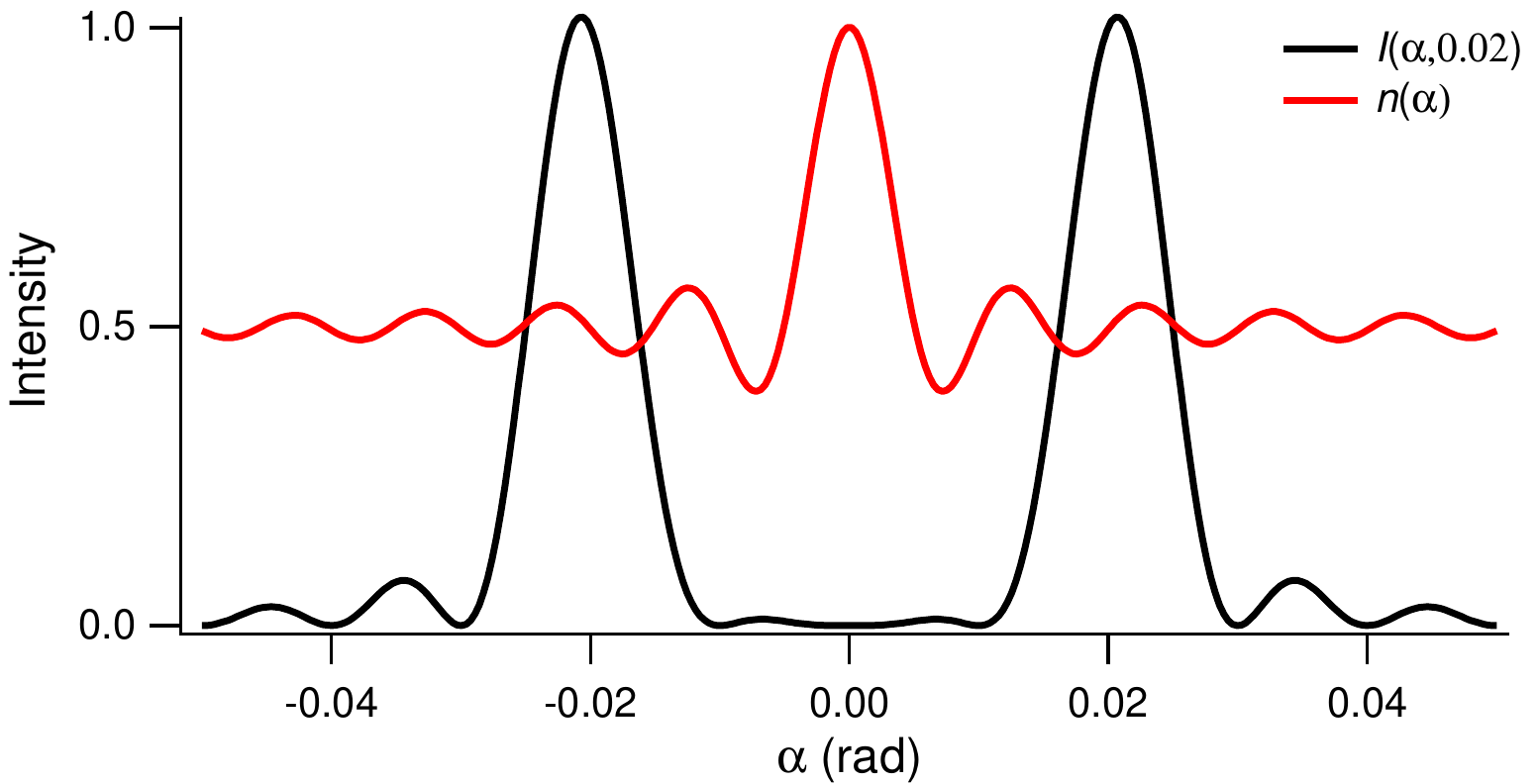}
\caption{
Two-slit-like interference pattern models the photoelectron intensity for coherent emission from $k_\omega$ and $-k_\omega$ states: $I(\alpha,\theta \propto k_\omega)$. The sum of such distributions for all pairs of states, $n(\alpha)$, models the momentum distribution function (see text).
\label{intf}}
\end{center}
\end{figure}

In Fig.\,\ref{intf} we show both, $I(\alpha,\theta)$ for $\theta = 0.02$ rad, and a function
\begin{eqnarray}
n(\alpha) = \int_{-\infty}^{\infty} I(\alpha,\theta) d\theta
\end{eqnarray}
that is analogous to the momentum distribution function discussed above. $E_k$ = 15.5 eV and $\xi = 100\pi$ that corresponds to $\lambda$ = 3 {\AA} and $d$ = 300 {\AA}. The parameter $\xi$ has been chosen to reproduce the width of the experimental MDCs.

As one can see, the simple interference model can explain both the appearance of the $k = 0$ peak in $n(k)$ and the fact that it looks narrower than the MDC peaks. So, although the photoemission from $k_\omega$ and $-k_\omega$ states is usually considered as non-coherent, one may assume that a finite electron-electron interaction can set the required coherence while the extended nature of topological surface states \cite{Seo}, due to very long phase-breaking lengths \cite{Checkelsky}, may provide a sufficient area on the sample surface from which the coherent photoemission occurs and results in the observed effect. The coherence of the $k_\omega$ and $-k_\omega$ states can be a consequence of a coherent spin rotation, recently suggested to explain the interference of spin states in photoemission from the surface alloy Sb/Ag(111) \cite{HugoDil}. Among other examples of interference in photoemission, that looks similar to our case, we want to mention the coherent photoemission from dimer molecule \cite{Hunt}, or the spin-state interference, observed in resonant photoemission induced by circularly polarized light from magnetized Gd \cite{Muller}.

\section{Conclusions}
We have found the profound deviations of ARPES spectrum of the topological surface states of Bi$_2$Se$_3$ from a generally expected spectral weight distribution of weakly interacting quasiparticles. The observed spectral line asymmetry can still be described by the one-particle spectral function but with the self-energy that strongly depends on both energy and momentum. The $\mathbf{k}\omega$-dependent constituent of the imaginary part of the self-energy that changes sign when crossing both the dispersion curves and the energy of the Dirac point can be a consequence a spin-flip scattering \cite{Yang}. In opposite, the the anomalously enhanced photoemission from the Dirac point is unlikely to be a self-energy effect if the Kramers-Kronig consistency is taken into account. In this situation a final state interference looks more plausible explanation for it.

\appendix

\begin{figure*}
\begin{center}
\includegraphics[width=0.9\textwidth]{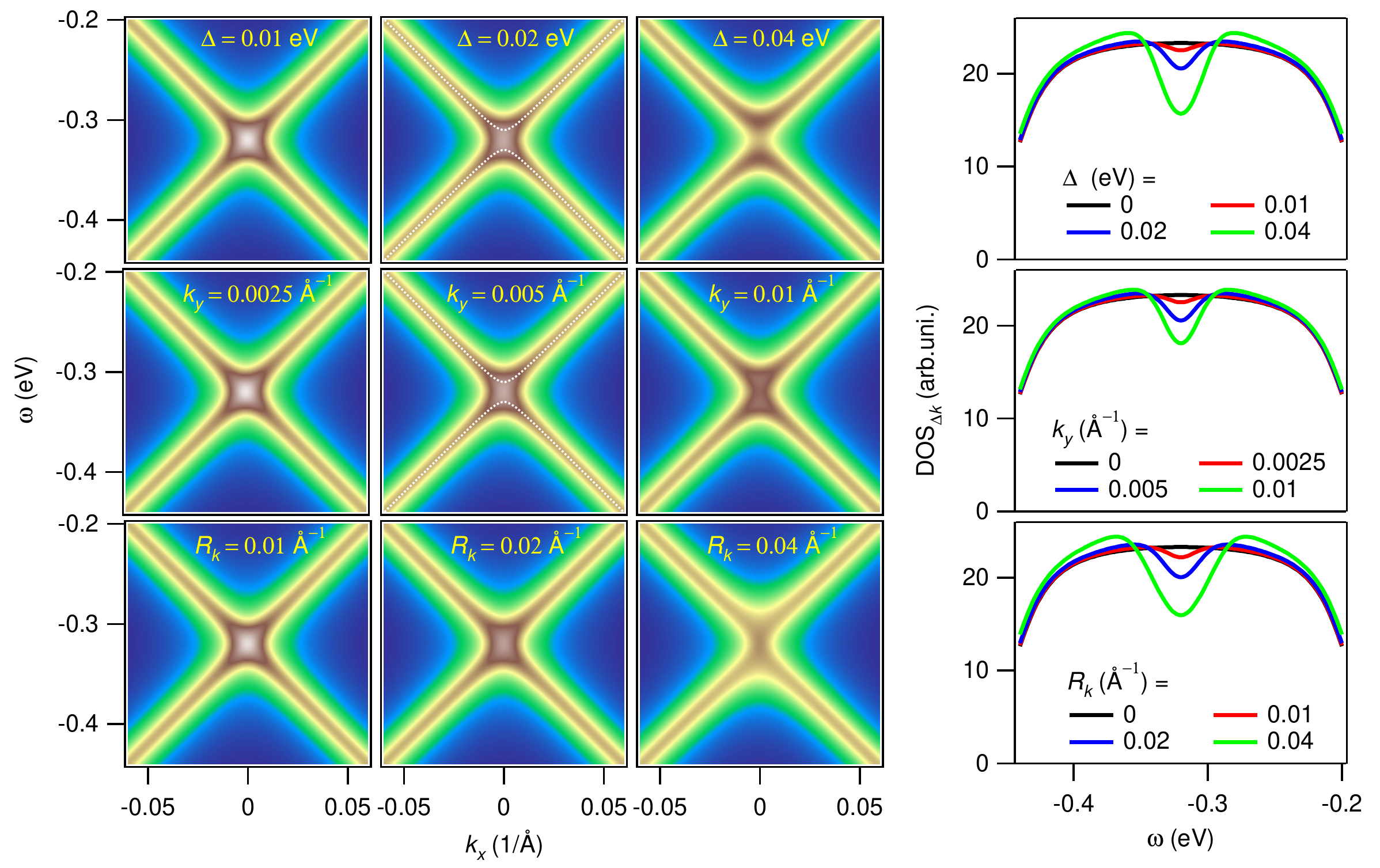}
\caption{
Simulation of a finite band gap (top row), not exact position in $\mathbf{k}$-space (middle row), and a finite momentum resolution (bottom row). All images are shown in the same absolute color scale. The dispersion, if shown, is presented by a white dotted line over corresponding images.
\label{Gap}}
\end{center}
\end{figure*}

\begin{figure*}[t]
\begin{center}
\includegraphics[width=\textwidth]{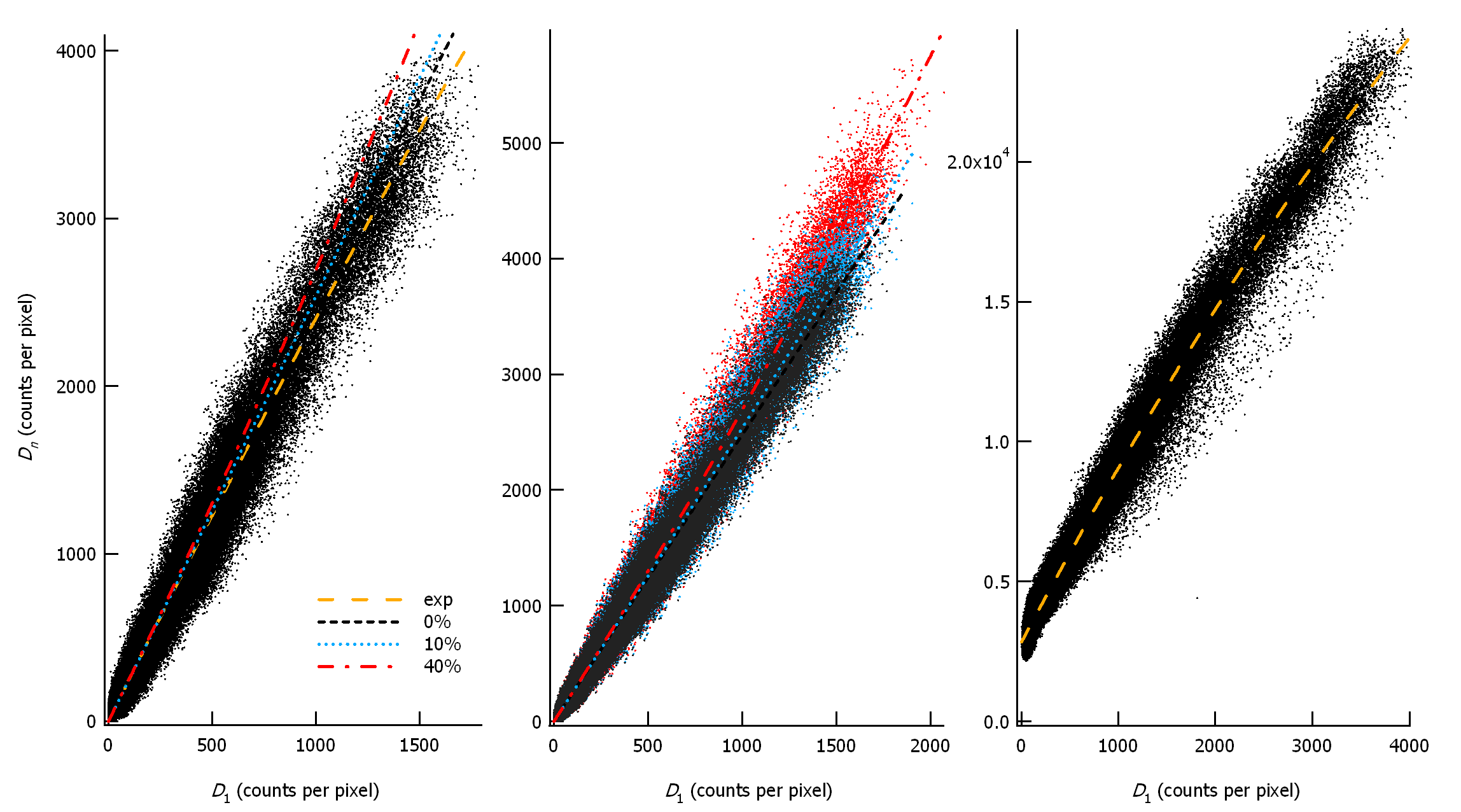}
\caption{
The intensity of each pixel of one ARPES spectrum, $D_n(\omega_i,k_j)$, \emph{vs} the intensity of the corresponding pixel of another, $D_1$, spectrum: spectra \#2 \emph{vs} \#1 (left); \#3 \emph{vs} \#2 (right); and simulated based on spectrum \#1 with $n = 2.5$ and $\alpha$ = 0, 0.1, and 0.4 (center), as described in the text.
\label{Detector}}
\end{center}
\end{figure*}

\section{\label{AppA}Effect of finite band gap and momentum resolution}

\noindent In Fig.\,\ref{Gap} we examine the effects of not exact position in $\mathbf{k}$-space and a finite momentum resolution, comparing them to the case of a finite band gap $\Delta$ due to lifting of the degeneracy at the Dirac point. We model the data as
\begin{eqnarray}
\begin{split}
I_\mathrm{mod}(\omega,k_x) = \frac{1}{R_k}\int_{-R_k/2}^{R_k/2}\left[\frac{\Sigma''}{(\omega - \varepsilon_+^\Delta)^2+{\Sigma''}^2}\right.+\\
\left.\frac{\Sigma''}{(\omega - \varepsilon_-^\Delta)^2+{\Sigma''}^2}\right] dk_y,
\end{split}
\label{A}
\end{eqnarray}
where $R_k$ is the momentum resolution perpendicular to the slit of analyzer, i.e. along $k_y$,
\begin{eqnarray}
\varepsilon_\pm^\Delta &=& \frac{\varepsilon_+ + \varepsilon_-}{2} \pm \sqrt{\left(\frac{\varepsilon_+ - \varepsilon_-}{2}\right)^2 + \frac{\Delta}{2}},\\
\varepsilon_\pm &=& \pm v_F \sqrt{k_x^2+k_y^2} - \omega_\mathrm{DP},
\end{eqnarray}
$v_F$ is the bare band velocity, $\omega_\mathrm{DP}$ is the energy of the Dirac point. Here we use $v_F$ = 2 eV{\AA}, $\omega_\mathrm{DP}$ = 0.32 eV, and $\Sigma''$ = 0.02 eV, to be close to the experimental data.

The images in Fig.\,\ref{Gap} represent the examples of $I_\mathrm{mod}(\omega,k_x)$ calculated for different values of each of the parameters $\Delta$, $k_y$, or $R_k$, while others where kept zero. One can see all these cases result in a dip rather than a peak in DOS$_{\Delta k}(\omega)$ and, by definition, do not effect $n_{\Delta\omega}(k)$. This result is natural since the dispersion at finite $k_y$ is similar to the dispersion with finite $\Delta$, which is known (e.g., from tunneling spectroscopy) to result in a dip in DOS. Consequently, at finite momentum resolution, one cannot expect different shape of DOS, which is just an average of partial DOSs with different $k_y$.

\section{\label{AppB}Detector linearity}

\noindent In order to estimate the linearity of the detector we compare pairs of standard ARPES images, such as shown in Fig.\,1, measured with different photon flux. Let us express the read-out intensity $D$ as a function of real intensity $I$ in each detector channel as $D^{i,j} = D(I_{i,j})$. Approximating the detector function as $D(I) \approx I + aI^2$, and assuming that real photoemission intensity can be changed from $I$ to $nI$ by changing the photon flux without changing other parameters of the light (such as energy, polarization, etc.), for two spectra measured with different photon flux one can write
\begin{eqnarray}
D_1 &=& D(I) \approx I (1 + aI),\\
D_n &=& D(nI) \approx nI (1 + anI).
\end{eqnarray}
Then, for small $aI$,
\begin{eqnarray}\label{DnvsD1}
D_n \approx nD_1 + a n(n-1) D_1^2,
\end{eqnarray}
so, one can estimate both $n$ and $a$ comparing two spectra measured with different photon flux.

In Fig.\,\ref{Detector}, we plot $D_n^{i,j}$ \emph{vs} $D_1^{i,j}$, i.e., the intensity of each pixel of one ARPES spectrum \emph{vs} the intensity of the corresponding pixel of another spectrum. Here we use the data for three spectra: \#1, \#2, and \#3, that differ by size of the photon beam from the undulator (controlled by the undulator blades: 0.5$\times$0.5, 0.8$\times$0.8, and 4$\times$4 mm$^2$, respectively). So, we plot \#2 vs \#1 data in Fig.\,\ref{Detector} (left panel) and \#3 vs \#2 data in Fig.\,\ref{Detector} (right panel). The data are shown in counts per pixel that is 1.5\,meV$\times$0.02$^\circ$ of size and is one of 294192 (454$\times$648) pixels in each image. For the recorded spectra the time per spectral channel was 453\,s, so the maximum counts recorded per pixel, $I_{max}$ = 6$\cdot$10$^3$ cpp, correspond to the count rate 13 counts per pixel per second, or to 0.43 counts per second per 1 $\mu \mathrm{m}^2$ of the detector area. The fit of the data in Fig.\,\ref{Detector} (left panel) to Eq.\,(\ref{DnvsD1}) gives $n = 2.509\pm0.002$ and negative $a = -1.04\times10^{-4}\pm1.5\times10^{-6}$ cpp$^{-1}$, or, in a dimensionless form, $\alpha = aI_{max} = -0.165\pm0.002$. For comparison, we add three lines that represent $D_n(D_1)$ function for different $\alpha$ = 0, 0.1, and 0.4. Those values are chosen to be close to the experimentally measured non-linearity of different detectors: 10\% for SPECS Phoibos 150 \cite{Wicks}, and 40\% for the Scienta SES200 detectors \cite{Kay,Mannella}. The fit of the data for higher count rates (Fig.\,\ref{Detector} right panel) gives $n = 6.5$ and $\alpha = -0.05$.

Since such a procedure of the evaluation of the detector linearity is not widely used, we demonstrate its applicability in Fig.\,\ref{Detector} (central panel). We model both $D_1$ and $D_n$ based on the spectrum \#1, taking the experimental data as $I$ and calculating $D_1^{i,j} = D(I_{i,j})$ and $D_n^{i,j} = D(nI_{i+1,j+1})$ with $n = 2.5$ and different $\alpha$ = 0, 0.1, and 0.4. The one-pixel shift of $D_n$ is introduced to simulate the experimental noise. The fits of the corresponding distributions to Eq.\,(\ref{DnvsD1}), shown for each distribution, are also shown in the left panel for comparison to the real data.

This analysis allows us to conclude that the linearity of the new Scienta detectors is significantly improved: the non-linearity of the detector we use is of the opposite sign but much smaller in absolute value than that has been reported for the old detectors \cite{Kay,Mannella}.

\section*{Acknowledgements}
We acknowledge discussions with E. E. Krasovskii, A. N. Yaresko, and A. V. Chubukov. The project was supported by the DFG under Grants No. KN393/4, BO 1912/2-1, and priority programme SPP 1458.


\onecolumngrid
\end{document}